\documentclass[conference]{IEEEtran}
\IEEEoverridecommandlockouts

\usepackage{cite}
\usepackage{algorithm}
\usepackage{algorithmic}

\ifCLASSINFOpdf
  \usepackage[pdftex]{graphicx}
\else
\fi

\usepackage{amssymb}
\usepackage{xcolor}
\usepackage{amsmath}
\usepackage{epsfig}
\usepackage{comment}
\usepackage{subfigure}
\usepackage{subcaption}
\usepackage{multirow}
\usepackage{makecell}
\usepackage{array}
\usepackage[lofdepth,lotdepth]{subfig}
\usepackage{subcaption}
\usepackage{graphicx}
\usepackage{flushend}

\begin{document}

\title{Antenna Tilt Failure Detection and Estimation via Integrated Sensing and Communications}

\author{\IEEEauthorblockN{Samed KEŞİR\IEEEauthorrefmark{1}\IEEEauthorrefmark{2}, Batuhan KAPLAN\IEEEauthorrefmark{1}\IEEEauthorrefmark{3}, Emre ARSLAN\IEEEauthorrefmark{1}, Ahmet Faruk COŞKUN\IEEEauthorrefmark{1}\\
\IEEEauthorblockA{\IEEEauthorrefmark{1}6GEN Laboratory, Next-Generation R\&D, Network Technologies, Turkcell, {\.{I}}stanbul, Turkiye}
\IEEEauthorblockA{\IEEEauthorrefmark{2}Electrical \& Electronics Engineering, Boğaziçi University, İstanbul, Turkiye}
\IEEEauthorblockA{\IEEEauthorrefmark{3}Electronics \& Communications Engineering, İstanbul Technical University, İstanbul, Turkiye}
{Emails: {\{samed.kesir, batuhan.kaplan, emre.arslan, coskun.ahmet\}@turkcell.com.tr,}}}}

\maketitle

\begin{abstract}

This paper addresses the critical sensitivity issue of narrow-beam communication systems to physical misalignments and exploits the potential of Integrated Sensing and Communications (ISAC) technology to propose a sensor-free antenna tilt failure detection and estimation framework. The proposed methods utilize environmental static clutter as geometric anchors to monitor systematic gain shifts in clutter heat maps. The proposed methods are introduced for precise antenna tilt detection and estimation using the standard 5G NR frame structure and two different waveforms. Numerical results show the potential of the proposed framework to enable autonomous, self healing network maintenance without the need for external sensors.

\end{abstract}


\section{Introduction}
Next-generation wireless networks target zero-touch management to enable self-configuration, and healing with minimal human intervention \cite{ETSI}. In this context, integrated sensing and communications (ISAC) is regarded as a key enabler and a cornerstone of 3GPP Release~19 discussions \cite{3GPPISAC,isac}, allowing base stations (BSs) to function as sensors for monitoring both the environment and network infrastructure integrity.

Cellular networks are shifting to higher frequency bands to meet capacity demands, however, to overcome the resulting severe path loss, next-generation BSs rely on narrow, highly directional beams\cite{Intro3,Intro4}. However, this reliance significantly increases sensitivity to antenna misalignment, leading to a rapid degradation in system tolerance.
Mechanical misalignments are inevitable in outdoor environments due to both short-term (e.g., natural disasters, storms) and long-term (e.g., thermal expansion, structural vibrations, wind, and corrosion) effects. Even small antenna tilt deviations can cause severe performance degradation or beam mispointing, leading to coverage holes or increased interference and, consequently, critical network reliability failures \cite{Intro5}. Despite their impact, existing fault detection methods are costly and unscalable, relying on external sensors or physical site inspections, and are largely reactive, requiring manual intervention after significant performance degradation \cite{Intro6}. This leads to prolonged suboptimal coverage and exposes personnel to hazardous conditions during tower maintenance. Although proactive hardware-based solutions (e.g., external sensors or gyroscopes) exist, they are economically infeasible and difficult to scale for ultra-dense 6G deployments. Consequently, a remote, autonomous, and hardware-free mechanism that validates antenna orientation using the existing radio infrastructure is highly attractive.

This study utilizes the potential of ISAC to enable a hard-ware free antenna tilt detection framework supporting self-healing and autonomous networks. 
Current ISAC-based calibration methods primarily focus on beam alignment with mobile users or cooperation with reconfigurable intelligent surface technology \cite{Intro7}. However, to the best of our knowledge, this work is the first to propose a unique approach to enable a hardware-free and tilt detection framework utilizing environmental scatterers. Unlike conventional approaches, our study exploits the BS’s sensing capability to utilize the surrounding static environment (i.e. buildings, towers, and poles) as geometric anchors. The core principle relies on the systematic shift of the sensed environment (i.e. generated radar image/point-cloud) correlating to an angular deviation in the antenna’s physical orientation. By periodically comparing the real-time sensing data with respect to a calibrated reference fingerprint, the network can autonomously detect antenna tilt failures (ATFs). Through this approach the BS or network node transforms 
into a self-aware system capable of proactive diagnosis, fulfilling the vision of a fully autonomous, self-healing next-generation radio environment.

The remainder of this paper is organized as follows: Section II presents the frame-structure and waveforms enabling ISAC while Section III discusses clutter heat maps which are the data sets utilized for ATF examination. Section IV dives into the analysis of ATF detection and estimation using the generated datasets. Numerical results are discussed in Section V and finally the paper is concluded in Section VI.

\section{ISAC-Enabling Frame Structure and Waveforms}
This section introduces the frame structure, signal models, and radar-like sensing mechanism that enables a cellular base station to acquire ISAC capabilities by leveraging the 5G NR frame structure, allowing shared utilisation of the frame resources for both communication and sensing functions.
\begin{figure}[t]
    \centering
    \includegraphics[width=0.98\linewidth]{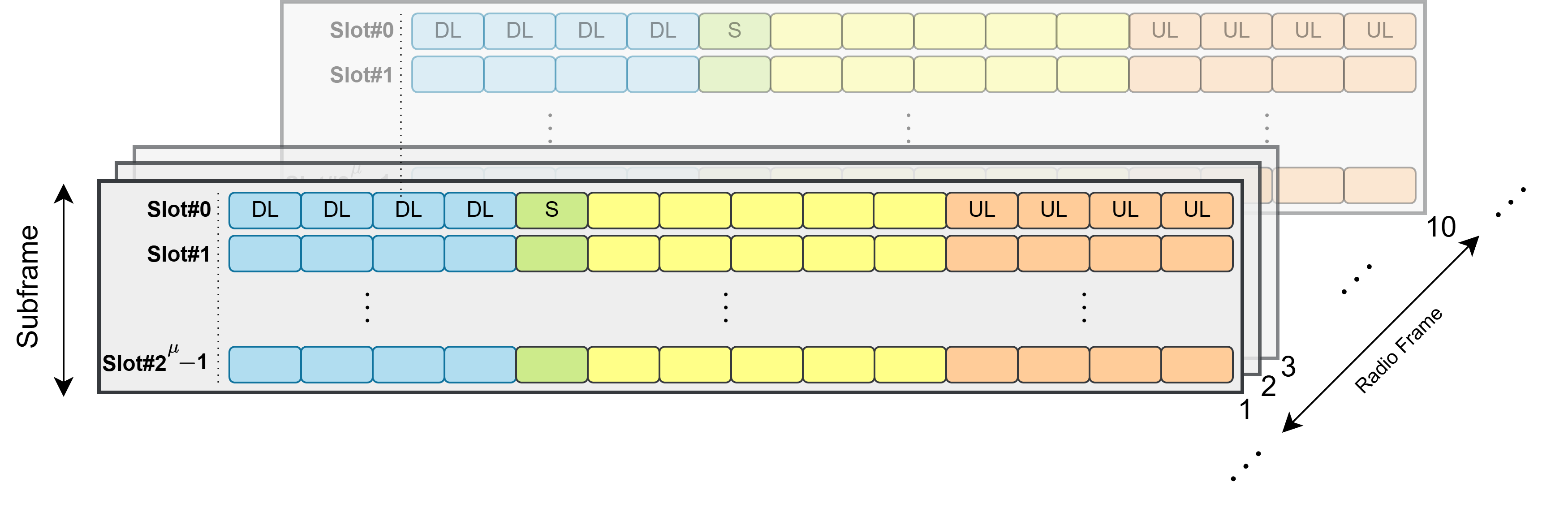}
    \caption{ISAC capable 5G NR frame configuration.}
    \label{fig:Figure1}
\end{figure}

\subsection{5G NR-Compliant Slot Formation}
5G NR employs a scalable numerology architecture defined by $\mu \in \{0, 1, \dots, 6\}$, where each subframe is further subdivided into $2^{\mu}$ slots. The OFDM symbols within each slot are strategically partitioned to support the downlink (DL), uplink (UL), and sensing (S) utilization. Fig. \ref{fig:Figure1} exhibits the 5G NR frame structure and illustrates the integration of the sensing signal. Considering the normal cyclic-prefix (CP) mode in 5G NR, each slot consists of $14$ OFDM symbols, where the symbol duration is determined by subcarrier spacing (SCS). 

The slot structure illustrated in Fig. \ref{fig:Figure1} can be configured in two different ways depending on whether the sensing waveform is based on OFDM or on linear frequency modulation (LFM) \cite{Skolnik2008}, which is widely adopted in traditional radar systems due to its distinct autocorrelation properties. In the first configuration, where the OFDM waveform is utilised, the first $N_D$ OFDM symbol durations within a slot are allocated to downlink communication. The last of these symbols is additionally reused as a reference sensing waveform to illuminate the environment according to radar principles. As a result, this configuration does not require additional time resources for the sensing reference waveform. In this case, radar echo reception is performed over the subsequent $N_{SR}$ symbol durations, during which the received signal samples are forwarded to the sensing processing blocks{\textcolor{black}{, with no communication transmission}}. Following the echo reception interval, the final $N_U$ symbol durations of the slot are allocated for uplink communication. In contrast, when the LFM waveform is selected, the number of downlink OFDM symbols is reduced by one compared to the OFDM-based configuration, and the LFM waveform is transmitted over a single symbol duration. Consequently, the proposed ISAC structure exhibits a trade-off between downlink data rate and enhanced sensing accuracy across the two configurations.

{\textcolor{black}{Note that the considered slot formation does not ensure an optimal communication–sensing trade-off, but the proposed approaches remain valid as long as the resulting clutter heat maps exhibit similar structures for different slot and/or resource grid configurations.}}

\subsection{OFDM and Chirp-Based Sensing Waveforms}
Two distinct signal waveform strategies are used depending on the communication and sensing requirements. For the OFDM-based sensing approach, the waveform corresponding to the transmission interval might be written as
\begin{equation}
     s_T(t) = \sum_{k=0}^{N_{SC}-1} d_{k}\exp\left({j 2\pi k \Delta f t}\right), \quad 0\le t\le \Delta T_s
\end{equation}

\noindent where $d_{k}$ represents the complex data symbol mapped to the $k$-th subcarrier, $N_{SC}$ is the number of total subcarriers, $\Delta T_s$ is the symbol duration, and $\Delta f$ defines the SCS calculated as $2^\mu\times 15 \text{ kHz}$ based on the specified numerology. For the LFM-based sensing approach, the emitted waveform within the reserved one-symbol interval might be expressed as \cite{Skolnik2008}
\begin{equation}
    s_T(t) = \exp\left({j\pi\frac{BW}{\Delta T_s}t(t-\Delta T_s)}\right),  \quad   0\le t\le \Delta T_s
\end{equation}
\noindent where $BW$ represents the occupied bandwidth.

\subsection{Radar-like Sensing Mechanism}
With the defined slot formation, conventional radar-based signal processing can be performed by computing the cross-correlation between the reference sensing waveform $s_T(t)$, which is transmitted in either OFDM or LFM form, and the signal samples collected during the reception interval \cite{Skolnik2008}
\begin{equation}
    r(\tau) = \int_{0}^{\Delta T_s} s_T(t) s_{R}^*(t + \tau) dt, \quad \tau \in \left[0,N_{SR}\Delta T_s\right]
\end{equation}
where $s_R(t)$ denotes the superposed radar echo that might be expressed as 
\begin{equation}
    s_R(t) = \sum_{k=1}^{K} \beta_k G^2(\theta_k, \phi_k) s_{T}(t - \tau_k) e^{j 2\pi f_{D,k} t} + n(t).
    \label{eq:received_signal}
\end{equation}
Here, $K$ denotes the number of scatterers, while $\beta_k$, $\tau_k$, and $f_{D,k}$ represent the complex channel coefficient (including RCS and path loss), time delay, and Doppler shift of the $k$-th target, respectively. $G(\theta_k,\phi_k)$ is the 3GPP-based base station antenna gain \cite{3GPPPattern}, squared to account for two-way propagation, and $n(t)$ denotes additive white Gaussian noise (AWGN) with variance $N_0 = kT_{\text{sys}}BWF_n$.

\section{Data Set: Clutter Heat Maps}
This section describes the generation of clutter heat maps (CHMs), which form the dataset used by the proposed ATF detection and estimation algorithms. Widely adopted in radar sensing \cite{CHM}, CHMs characterize signal power over domain-dependent parameters. In this study, CHMs represent the average radar echo power over range and azimuth within the sensing service area of the considered ISAC scenario, following an environment-mapping approach. Each CHM entry is updated via the cross-correlation defined in (3). The service area is specified by a maximum sensing range $R_{\max}$ and an azimuth sector width $\Delta\phi$, where $R_{\max} = c N_{SR}\Delta T_s/2$ is determined by the reception duration allocated after sensing waveform transmission in the 5G NR slot. The azimuth sector is discretized into $N_{\phi}$ bins with $1^{\circ}$ resolution. For each azimuth direction, $N_{\mathrm{sl}}$ slot sequences are transmitted according to the 5G NR slot configuration, and the collected echoes yield range profiles (RPs) dependent on target RCS and mobility.

The $N_{\mathrm{sl}}$ radar echoes collected per azimuth can be processed via Fourier-based methods to obtain range–Doppler characteristics or averaged to generate CHMs. In this study, CHMs serve as the primary data source and represent environment-mapping outputs. By sequentially scanning each azimuth bin, the entire sector of interest is covered over $N_{\phi}$ frame durations using the sensing resources defined by the slot structure. The scan cycle is repeated to enable continuous sensing for target detection and tracking, as widely studied in the 5G ISAC literature \cite{5GISACLit}. 
Within this framework, CHMs{\footnote{The ATF-oriented clutter heat maps have been uploaded to IEEE DataPort as a standard dataset (https://dx.doi.org/10.21227/x6pp-z729).}, which are used for moving and stationary target detection as well as environment mapping, are updated at each scan cycle using a sliding window of $N_t$ radar scans. To this end, a recursive moving average (RMA) filter is applied to stabilize the CHM entries $y_k$ after the $k$th scan, defined as:
\begin{equation}
y_{k} = \alpha y_{k-1} + (1-\alpha)x_k,
\label{eq:rma_filter}
\end{equation}
where $y_k$ is the smoothed output, $x_k$ is the instantaneous clutter power, and $\alpha\triangleq (N_t-1)/N_t$ is the forgetting factor.

\begin{figure*}[t]
    \centering
    \subfigure[Sensing area including the static and moving scatterer locations]
    {\includegraphics[width=0.62\columnwidth]{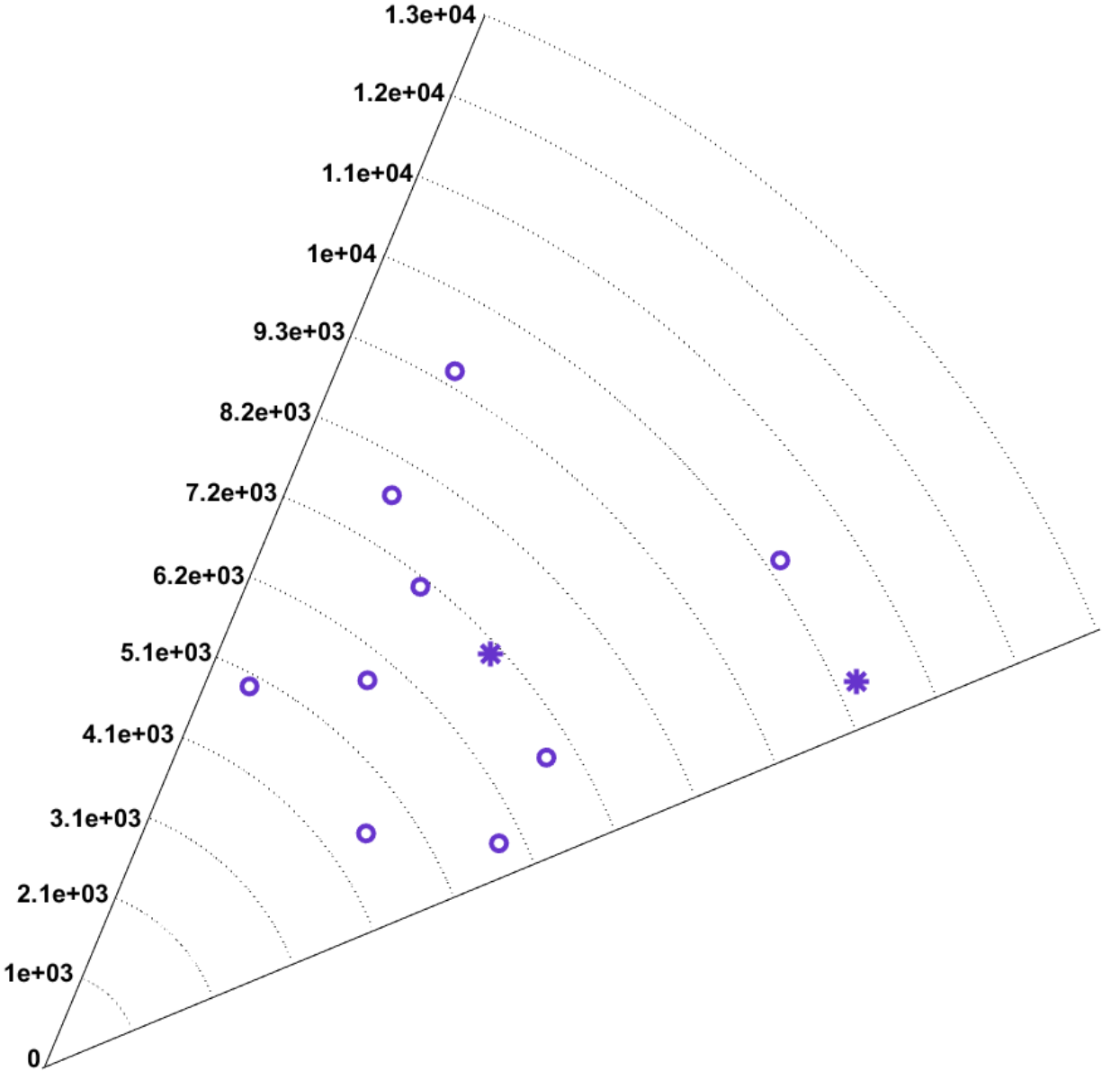}}
    \subfigure[Clutter heat map generated for OFDM-based waveform]
    {\includegraphics[width=0.62\columnwidth]{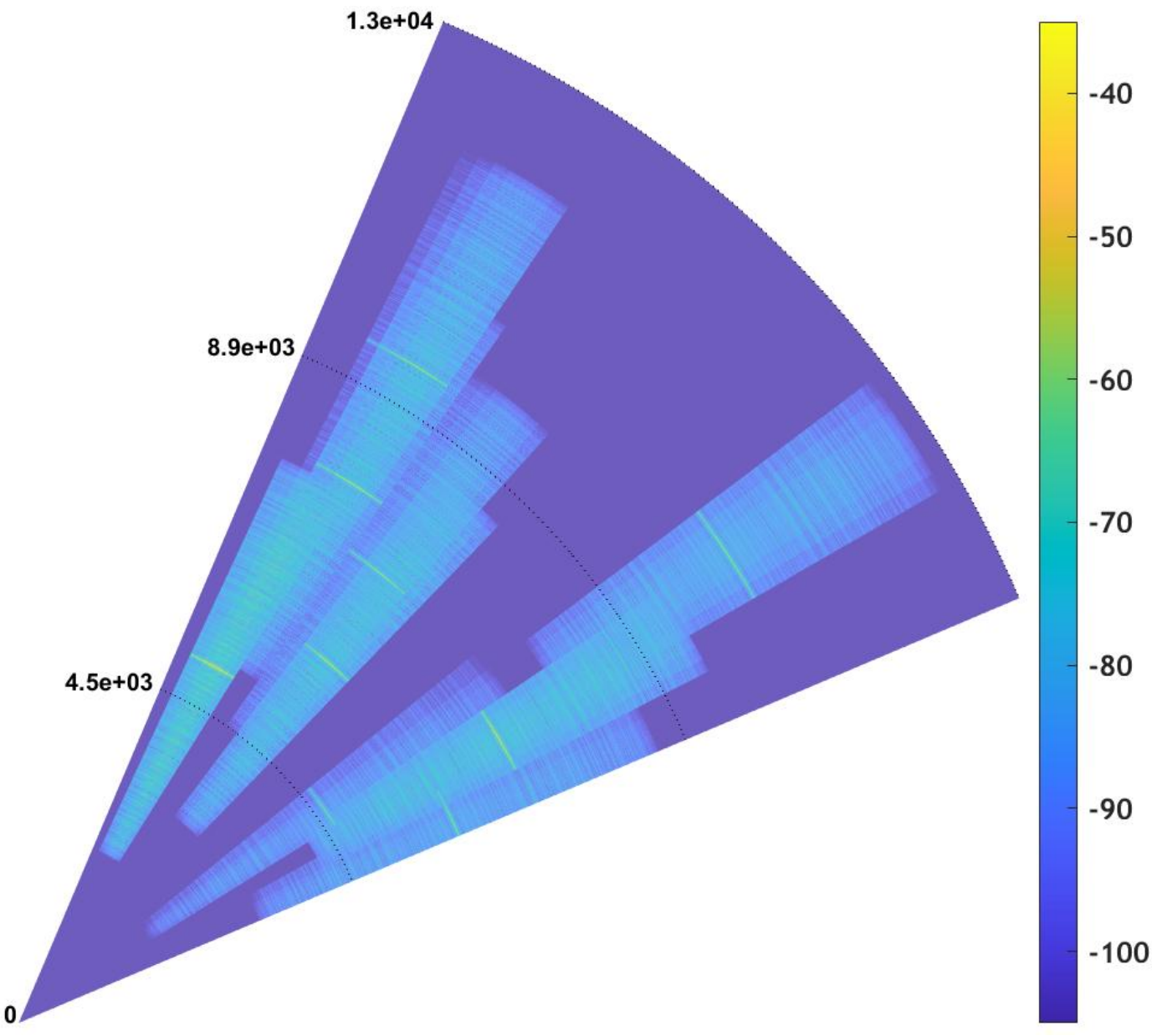}}
    \subfigure[Clutter heat map generated for LFM-based waveform]
    {\includegraphics[width=0.62\columnwidth]{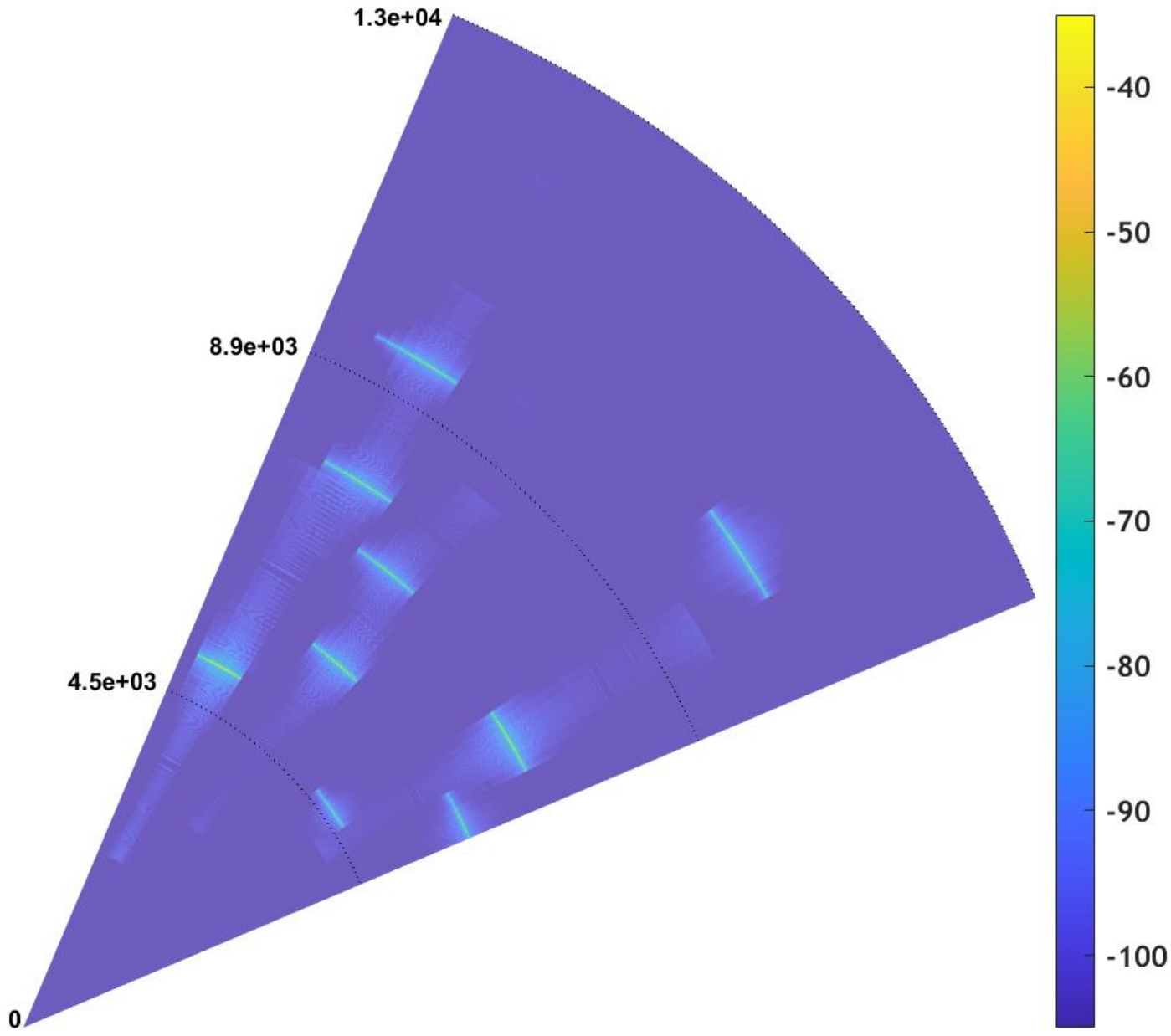}}
    \caption{Sensing area and corresponding clutter heat maps for two waveforms}
    \label{fig:Figure2}
\end{figure*}

\section{Antenna Tilt Failure Analysis}

This section details the methodology for analyzing ATFs in ground-based ISAC systems. The analysis framework comprises three stages: understanding the effect of tilt on clutter returns, real-time failure detection, and the estimation of antenna tilt misalignment.

\subsection{Effects of Tilt Misalignment on Sensing Performance}
In ground-based sensing scenarios, although the received power from static clutter exhibits stochastic variations, its dominant behavior is largely deterministic, driven by the radar equation and the BS antenna radiation pattern. For a static scatterer located at a radial distance $r$, the geometric elevation angle (measured from the zenith) is defined as $\theta_{\text{g}}(r) = (\pi/2) + \arctan\left({\Delta h}/{r}\right)$
where $\Delta h\triangleq h_{\text{BS}} - h_{s,r}$ is the difference between BS antenna height and the scatterer height $h_{s,r}$. Under normal operation with a mechanical downtilt $\theta_{o}$, the effective elevation angle relative to the antenna boresight is $\theta_{\text{eff}} = \theta_{\text{g}} - \theta_o$. A mechanical tilt failure introduces an unknown offset $\Delta_{\theta}$, shifting the effective angle to $\theta'_{\text{eff}} = \theta_{\text{g}} - (\theta_o + \Delta_{\theta})$. This angular shift results in a deviation in the two-way antenna gain, $G^2(\cdot)$, which results in considerable gain degradation in CHM entries. Since the environment is static, abrupt deviations in the CHM power levels are indicative of antenna orientation failures. {\textcolor{black}{Static scatterers serve as anchor points, since they are typically present and the averaging nature of the CHM suppresses mobility effects, target and UE mobility does not significantly affect the proposed approaches. In addition, although this work does not propose a method for static-moving target separation, the proposed detection and estimation algorithms can be enhanced with a preliminary separation step (e.g., focused on Doppler phase characteristics).}}
\subsection{Antenna Tilt Failure (ATF) Detection Algorithm}
To enable low-complexity system health monitoring, we adopt a lightweight detection algorithm that exploits the stability of azimuth-averaged CHM RPs. Let $\bar{P}_k(r)$ denote the azimuth-averaged received power profile as a function range $r$ for scan index $k$. At this stage, the CHM RPs prior to an ATF event are expected to exhibit a stable characteristic across consecutive scans, apart from minor fluctuations. Following the occurrence of ATF, this stability is anticipated to degrade to a discriminable level. To assess this stability, the proposed ATF detection algorithm monitors the linearly scaled magnitude-weighted ratios (MWR) of CHM RPs obtained from successive scans:
\begin{flalign}
\eta_k(r)
&=
\frac{\max\!\left(|\bar{P}_k(r)|,\,|\bar{P}_{k-1}(r)|\right)}
{\max\!\left(|\bar{P}_k(r)|,\,|\bar{P}_{k-1}(r)|\right)+\tau}
\left(
\frac{\bar{P}_k(r)}{\bar{P}_{k-1}(r)} - 1
\right) \nonumber \\
&\qquad \text{for } |\bar{P}_k(r)|>0 \;\wedge\; |\bar{P}_{k-1}(r)|>0
\end{flalign}
where $\tau > 0$ is the regularization parameter. The proposed MWR metric is inspired by magnitude-weighted comparison principles commonly used in radar signal processing and clutter analysis \cite{Skolnik2008}, while incorporating a numerical stability constant to avoid ill-conditioned ratio behavior \cite{Tikhonov1977}. A failure is flagged if the deviation $\eta_k(r)$ exceeds a predefined threshold $\gamma_{\text{det}}$ (e.g., $0.01$) for a statistically significant number of range bins. This metric is robust against AWGN but highly sensitive to the systematic gain shifts induced by tilt failures.

\subsection{ATF Estimation Algorithms}
We propose two numerical methods to address the challenge of estimating the unknown tilt offset ${\hat\Delta_{\theta}}$ despite the smoothing effect of the RMA filter.

\subsubsection{Method I: Instantaneous Pattern Matching}

This method is based on analyzing the difference between the incremental contributions $x_k$ and $x_{k-1}$ added to the updated CHM RPs over two consecutive scans, and associating this difference with a precomputed look-up table derived from the 3GPP antenna radiation pattern. To extract the marginal range-profile contributions from successive scans, CHM matrices from three consecutive scan cycles are required. Hence, this method is usable only when the scan index is greater than or equal to three. This limitation can be mitigated by setting the sliding-window length parameter introduced in Section III as $N_t \geq 3$.

To obtain the marginal CHM contributions, the RMA filter is mathematically inverted as $\hat{x}_k= \left({y_k - \alpha y_{k-1}}\right)/({1 - \alpha})$ for scan index $k$. Comparing the recovered instantaneous RPs $\hat{x}_k$ and $\hat{x}_{k-1}$, we compute the one-way gain reduction $\Delta G_{\text{dB}}(r)$ at each range bin. For each dominant peak at range $r$ (corresponding to elevation $\theta_{\text{g}}(r)$), the estimator finds the candidate tilt angle $\delta$ that minimizes the difference between the maximum pattern gain and the observed reduced gain:
\begin{equation}
    {\hat\Delta\phi} = \mathbb{E}_{r}[\arg \min_{\delta} \left| G_{{\text{dB}}}(\theta_{\text{g}}(r) - \delta) - (G_{\text{max}} - \Delta G_{\text{dB}}(r)) \right|].
    \label{eq:pattern_matching}
\end{equation}
This approach is computationally efficient as it relies on simple lookup operations after inverse filtering (Algorithm \ref{alg:pattern_match}).

\begin{algorithm}[!t]
\caption{Instantaneous Pattern Matching Estimation}
\label{alg:pattern_match}
\begin{algorithmic}[1]
\REQUIRE CHMs $\mathbf{Y}_{k}, \mathbf{Y}_{k-1}, \mathbf{Y}_{k-2}$, Window $W$, Radiation Pattern $G(\theta)$
\ENSURE Tilt Estimate $\hat\Delta_{\theta}$

\STATE $\alpha \gets (W-1)/W$ \COMMENT{Filter decay factor}
\STATE $\hat{\mathbf{X}}_{k-1} \gets (\mathbf{Y}_{k-1} - \alpha \mathbf{Y}_{k-2}) / (1-\alpha)$
\STATE $\hat{\mathbf{X}}_{k} \gets (\mathbf{Y}_{k} - \alpha \mathbf{Y}_{k-1}) / (1-\alpha)$
\STATE \emph{\% Compute one-way gain reduction (dB)}
\STATE $\Delta \mathbf{G} \gets |10\log_{10}(\hat{\mathbf{X}}_{k}) - 10\log_{10}(\hat{\mathbf{X}}_{k-1})| / 2$
\STATE $\mathcal{I} \gets \text{SelectPeaks}(\hat{\mathbf{X}}_{k-1})$
\STATE $\mathcal{D} \gets \text{PatternDatabase}(G)$
\FOR{each peak index $i \in \mathcal{I}$}
    \STATE $\theta_i \gets \text{GetElevation}(r_i)$
    \STATE $\hat{\delta}_i \gets \arg\min_{\delta} | \mathcal{D}(\theta_i - \delta) - (G_{\max} - \Delta \mathbf{G}[i]) |$
\ENDFOR
\RETURN $\hat\Delta_{\theta} \gets \text{Mean}(\hat{\delta})$

\end{algorithmic}
\end{algorithm}

\subsubsection{Method II: Transient Model Optimization}
This method utilizes the exact antenna pattern to model the transient response of the RMA filter. Instead of inverting the data, we define a cost function that predicts the \textit{expected} moving average output for a candidate tilt $\delta$. The theoretical instantaneous power ratio $\rho(\delta, r)$ for a candidate tilt is:
\begin{equation}
    \rho(\delta, r) = {|G(\theta_{\text{g}}(r) - \delta)|^2}/{|G(\theta_{\text{g}}(r))|^2}.
    \label{eq:theo_ratio}
\end{equation}
We model the filter's transient response to a step input, where the predicted power ratio $\hat{R}_{\text{pred}}$ at scanning instant $k$ is:
\begin{equation}
    \hat{R}_{\text{pred}}(\delta) = \rho(\delta) + \left[1 - \rho(\delta)\right]\alpha^{k}.
    \label{eq:transient_model}
\end{equation}
The optimal tilt is found by minimizing the error between the observed ratio $R_{\text{obs}}$ (from measured RPs) and the predicted transient ratio:
\begin{equation}
    \hat\Delta_{\theta} = \arg \min_{\delta} \sum_{r \in \mathcal{P}} \left(10\log_{10}\left({\frac{R_{\text{obs}}(r)}{\hat{R}_{\text{pred}}(\delta,r)}}\right)\right)^2,
    \label{eq:cost_function}
\end{equation}
where $\mathcal{P}$ is the set of peak range indices. This optimization uses a coarse grid search followed by local refinement to ensure global convergence (Algorithm \ref{alg:tilt_est}).

\begin{algorithm}[!t]
\caption{Tilt Estimation with Recursive Lag Modeling}
\label{alg:tilt_est}
\begin{algorithmic}[1]
\REQUIRE Range Profiles $\mathbf{P}$, Range $\mathbf{r}$, Window $W$, Fail Index $k_f$, Radiation Pattern $G(\theta)$
\ENSURE Estimated Physical Tilt $\hat\Delta_{\theta}$

\STATE $\alpha \leftarrow (W - 1) / W$ \COMMENT{Filter decay factor}
\STATE $\mathbf{p}_{\text{ref}} \leftarrow \text{Mean}(\mathbf{P}_{t < k_f})$ \COMMENT{Baseline profile}
\STATE $\mathcal{I} \leftarrow \text{SelectPeaks}(\mathbf{p}_{\text{ref}})$
\STATE $\mathbf{L}_{\text{obs}} \leftarrow 10\log_{10}(\mathbf{P}[\mathcal{I}, k_f] \oslash \mathbf{p}_{\text{ref}}[\mathcal{I}])$ \COMMENT{Observed log-ratio}
\STATE $\boldsymbol{\theta} \leftarrow \pi - \arctan(\mathbf{r}[\mathcal{I}] / h_{\text{BS}})$
\STATE $\mathbf{G}_{\text{ref}} \leftarrow |G(\boldsymbol{\theta})|^2$

\STATE \textbf{Optimization (Minimize Cost $J(\delta)$):}
\FOR{each candidate $\delta$ in search grid}
    \STATE $\boldsymbol{\theta}' \leftarrow \boldsymbol{\theta} - \delta$ \COMMENT{Apply tilt}
    \STATE $\mathbf{G}_{\text{tilt}} \leftarrow |G(\boldsymbol{\theta}')|^2$
    \STATE $\boldsymbol{\rho} \leftarrow \mathbf{G}_{\text{tilt}} \oslash \mathbf{G}_{\text{ref}}$ \COMMENT{Theoretical ratio}
    \STATE $\hat{\boldsymbol{\rho}} \leftarrow \boldsymbol{\rho} + (1 - \boldsymbol{\rho})\alpha$ \COMMENT{Model filter lag}
    \STATE $J(\delta) \leftarrow \|\mathbf{L}_{\text{obs}} - 10\log_{10}(\hat{\boldsymbol{\rho}})\|^2$
\ENDFOR

\STATE $\delta^* \leftarrow \arg\min_{\delta} J(\delta)$ \COMMENT{Grid Search + Refinement}
\RETURN $\delta^*$

\end{algorithmic}
\end{algorithm}

\section{Numerical Results}
This section evaluates the performance of the proposed ATF estimation methods using CHM data generated under a 5G NR–compliant ISAC configuration with two waveform types, focusing on fault detection and estimation. For ATF detection and estimation, the failure is assumed to occur after a fixed scan duration as a mechanically induced downward tilt increase, resulting in observable modifications in the CHM matrix in the subsequent scan cycle.

\subsection{Simulation Setup and Test Scenarios}
The 5G NR parameters used to generate the CHM datasets and the sensing search-space settings are listed in Table \ref{table1}.

\begin{table}[t]
  \centering
  \caption{\textsc{Simulation Parameters}}
  \label{table1}
  \begin{tabular}{c c c}
    \hline
    {\bf{Param.}} & {\bf{Definition}} & {\bf{Value}} \\
    \hline
    $f_c$ & Center frequency & $3700$ MHz \\
    $\mu$ & Numerology & $2$ \\
    $N_D$ & \#DL symbols (OFDM-based sensing) & $6$ \\
    $N_D$ & \#DL symbols (LFM-based sensing) & $5$ \\
    $N_{SR}$ & \#Symbols for sensing recept. & $5$ \\
    $N_U$ & \#UL symbols per slot & $3$ \\
    $BW$ & Signal bandwidth & $50$ MHz \\
    $P_t$ & Transmit Power & $1000$ Watts \\
    $N_h$ & \#Antenna elements in horizontal axis & $16$ \\
    $N_v$ & \#Antenna elements in vertical axis & $12$ \\
    $G_t(\theta,\phi)$ & 3GPP-defined radiation pattern & \cite{3GPPPattern} \\
    $\{\theta_{HP}$,$\phi_{HP}\}$ & Half-power beamwidths & $\{6^\circ,6^\circ\}$ \\
    $G_{max}$ & Maximum antenna gain in dB & $21$ \\
     & Pattern Front-Back Ratio in dB & $30$ \\
     & Side-Lobe Level in dB & $30$ \\
    BS$\left(x, y, z\right)$ & BS Location & $\left(0, 0, 60\right)$ \\
    $N_s$ & Number of sensing scans & $9$ \\
    $N_t$ & Number of training scans & $3$ \\
    $n_{\Delta\theta}$ & Sensing scan that hosts the ATF event & $5$ \\
    $\Delta\phi$ & Angular width of sensing RoI & $45^\circ$ \\
    $\phi_{B}$ & Azimuth bore-sight angle of BS & $45^\circ$ \\
    $N_{sl}$ & Number of slots per frame & $40$ \\
    $N_{\phi}$ & Number of azimuth slices per each scan & $45$ \\
    $F_{n,dB}$ & Noise figure at UE antenna & $5$ dB \\
    $T_{sys}$ & System temperature & $290$ K \\
    $\lambda$ & Path-loss exponent & $1.785$ \cite{PLERef}\\\hline
    \end{tabular}
\end{table}

Table \ref{table2} lists the range, azimuth, diameter, and height of the stationary and moving scatterers relative to the ISAC-capable base station at the coordinate origin, as illustrated in Fig. \ref{fig:Figure2}-(a). Stationary scatterers are modelled as cylindrical structures with predefined dimensions for simplified RCS modelling \cite{Balanis1989}, \cite{Coskun2015}. The two moving targets are assumed to have unit RCS and constant velocities of $\mathbf{v}_1 = [10.32; 6.14; 0]$ and $\mathbf{v}_2 = [-11.43; -3.68; 0]$, respectively. Fig. \ref{fig:Figure2}-(a) visualizes the nine stationary scatterers with circular markers and the two moving targets with star-shaped markers.

\begin{table}[t]
  \centering
  \caption{\textsc{Scatterer location information}}
  \label{table2}
  \begin{tabular}{c c c c c}
    \hline
    {\bf{Scatterer}} & {\bf{Range}} & {\bf{Azimuth}} & {\bf{Diameter}} & {\bf{Height}}\\
    {ID} & {[m]} & {$\circ$} & {[m]} & {[m]}\\
    \hline
    $1$ & $5003.4$ & $28.1$ & $32$ & $41$ \\
    $2$ & $5847.8$ & $39.9$ & $28$ & $32$ \\
    $3$ & $4594.1$ & $54.4$ & $15$ & $10$ \\
    $4$ & $6869.7$ & $58.7$ & $44$ & $50$ \\
    $5$ & $5891.4$ & $64.3$ & $20$ & $18$ \\
    $6$ & $7801.7$ & $31.2$ & $32$ & $38$ \\
    $7$ & $10460.2$ & $56.2$ & $40$ & $51$ \\
    $8$ & $7101.2$ & $38.1$ & $35$ & $27$ \\
    $9$ & $9443.6$ & $30.5$ & $36$ & $42$ \\
    $10$ & $7095.3$ & $47.3$ & - & - \\
    $11$ & $10509.2$ & $65.7$ & - & - \\\hline   
    \end{tabular}
\end{table}

\subsection{Visualization of processed CHMs for ATF Detection}
Using the parameters in Tables \ref{table1} and \ref{table2}, the CHM matrices generated for both waveforms are shown in polar form in Figs. \ref{fig:Figure2}-(b) and (c). To illustrate the effect of the ATF occurring at scan index $n_{s,a}=5$ and to provide intuition for the detection approach, the resulting CHM variations for the OFDM- and LFM-based waveforms are presented in Fig. \ref{fig:Figure3} and Fig. \ref{fig:Figure4}, respectively. The upper subfigures show CHM RPs obtained by azimuthal averaging for two pre-ATF scans (IDs $3$ and $4$) and the post-ATF scan. The pre-ATF profiles are nearly identical, whereas a clear deviation is observed in the RP following the ATF event (scan ID $5$).

\begin{figure}[t]
    \centering {\includegraphics[width=0.95\columnwidth]{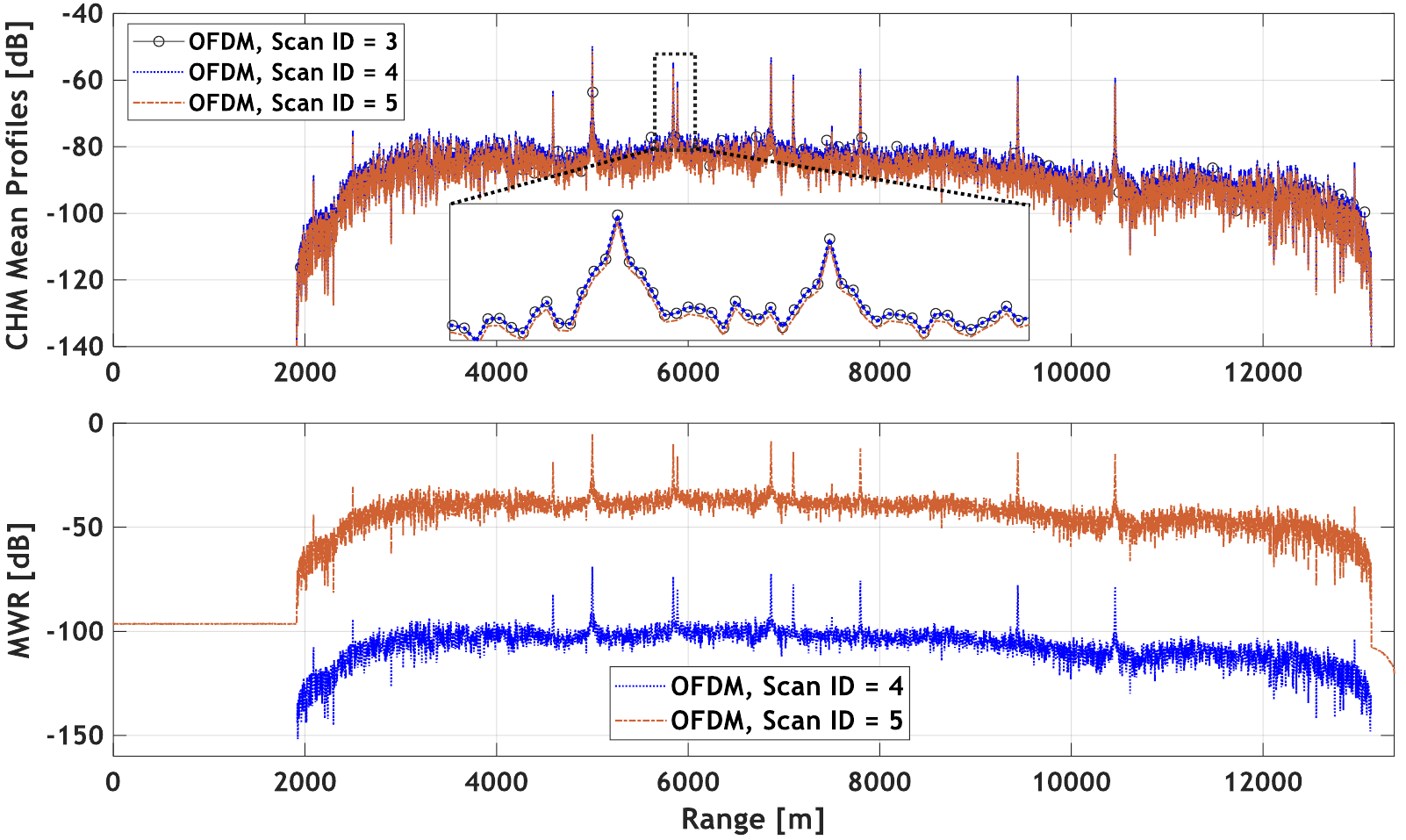}}
    \caption{The variation of averaged CHMs before and after ATF event: (upper) Averaged CHMs, (lower) Variation of weighted magnitude ratios (for OFDM-based echoes, and $\Delta_{\theta} = 4^\circ$)}
    \label{fig:Figure3}
\end{figure}

\begin{figure}[t]
    \centering {\includegraphics[width=0.95\columnwidth]{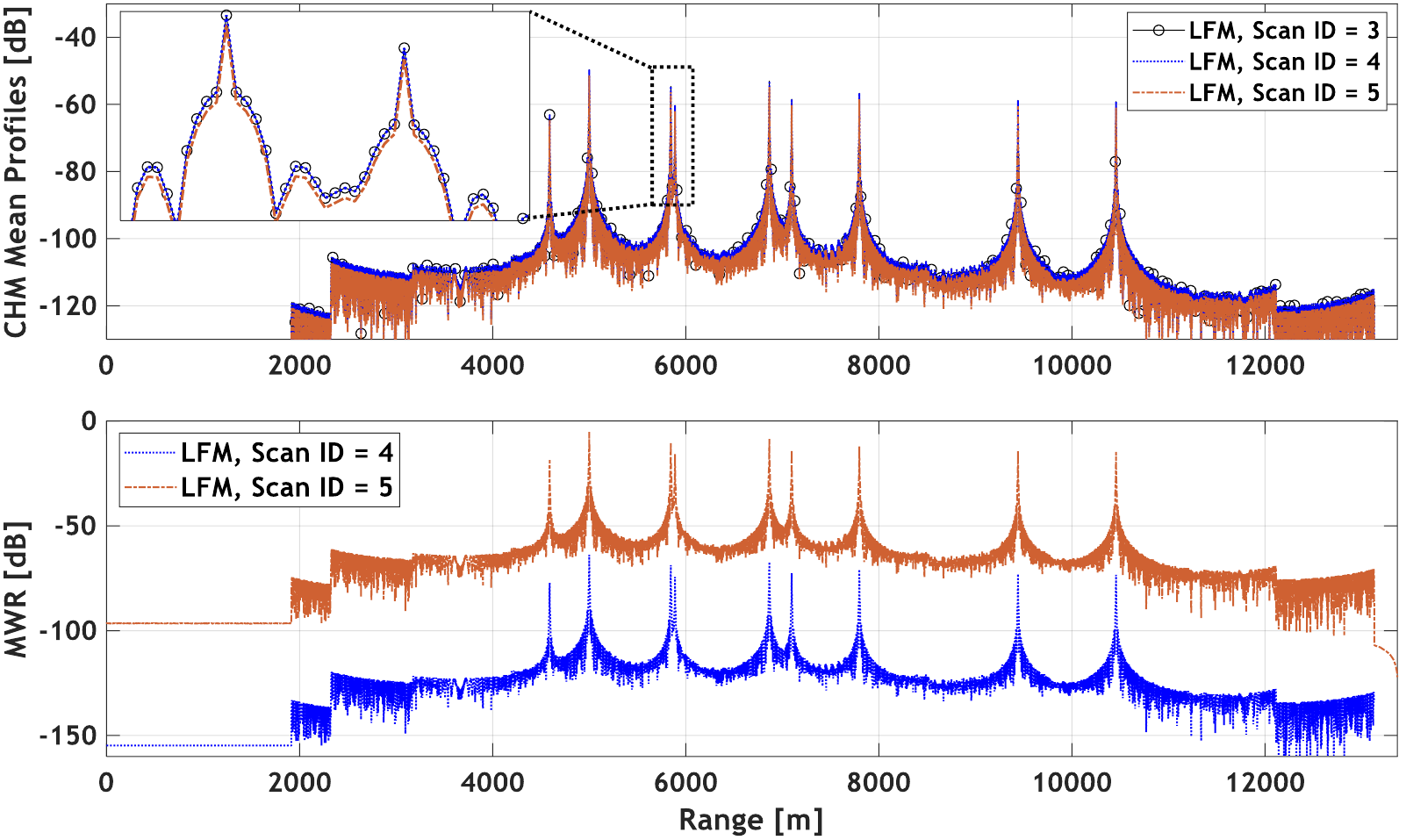}}
    \caption{The variation of averaged CHMs before and after ATF event: (upper) Averaged CHMs, (lower) Variation of weighted magnitude ratios (for LFM-based echoes, and $\Delta_{\theta} = 4^\circ$)}
    \label{fig:Figure4}
\end{figure}
To detect this divergence, the MWR variations defined in Section IV-B are computed at the end of the pre-ATF (scan ID $4$) and post-ATF (scan ID $5$) scans. The resulting variations, shown in the lower subfigures of Fig. \ref{fig:Figure3} and Fig. \ref{fig:Figure4} for both waveforms, demonstrate that the ATF event can be effectively detected using a threshold-based approach.

\subsection{Comparison on Utilized Estimation Methods}
After detecting the ATF event using the CHM matrices—the ISAC output—a performance comparison of ATF degradation estimation methods is performed based on CHM variations. The two methods introduced in Section IV-C are applied to CHM matrices generated with both waveforms, yielding scan-index-dependent estimates for $\Delta_{\theta} = {1.5, 3.25, 5}$ degrees, as shown in Fig. \ref{fig:Figure5}. For both waveforms, the estimation trends are similar across all degradation levels, while the two methods exhibit different error margins for small and large tilt errors.

\begin{figure}[t]
    \centering {\includegraphics[width=0.92\columnwidth]{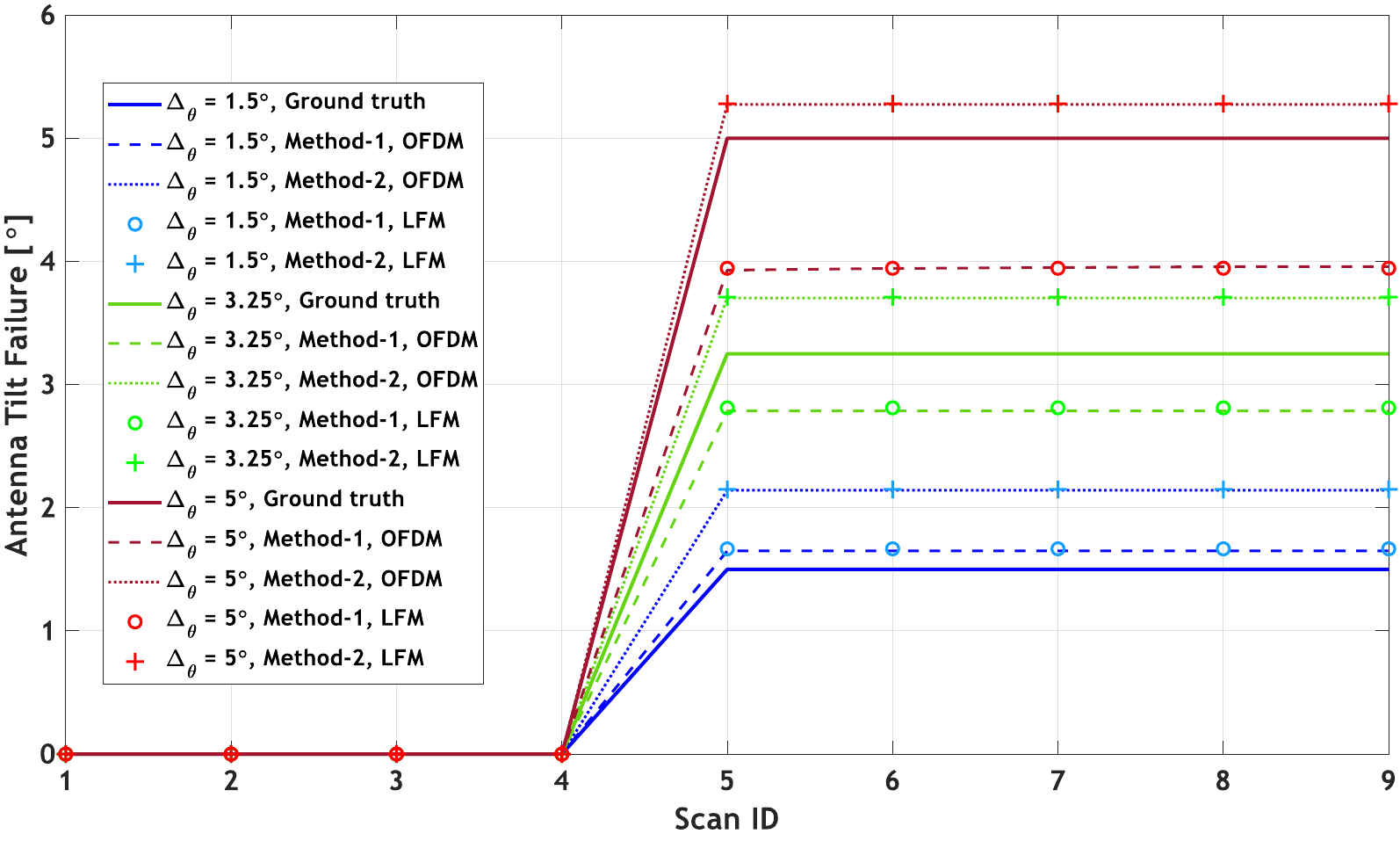}}
    \caption{Comparison of ATF estimation methods' outputs for selected ATF values}
    \label{fig:Figure5}
\end{figure}

\subsection{Extensive Examination on Estimation Accuracy}
To further characterise the estimation error as a function of ATF magnitude, additional simulations are conducted for $\Delta_{\theta}\in[0.5, 6]$ degrees. The method outputs are compared with ground truth, and the resulting absolute and relative errors are shown in Fig. \ref{fig:Figure6}. The bias and percentage error results (upper and lower subfigures) indicate consistent behaviour across both waveforms, while revealing distinct performance characteristics of the two methods. Method-2 exhibits decreasing errors for larger ATF values but higher errors than Method-1 in the low-degradation regime.
The complementary error trends suggest potential for accurate ATF estimation and, when needed, compensation via electronic tilt adjustment. However, the relatively high errors observed in certain ATF ranges motivate the development of more advanced estimation algorithms with improved performance.
\begin{figure}[t]
    \centering {\includegraphics[width=0.95\columnwidth]{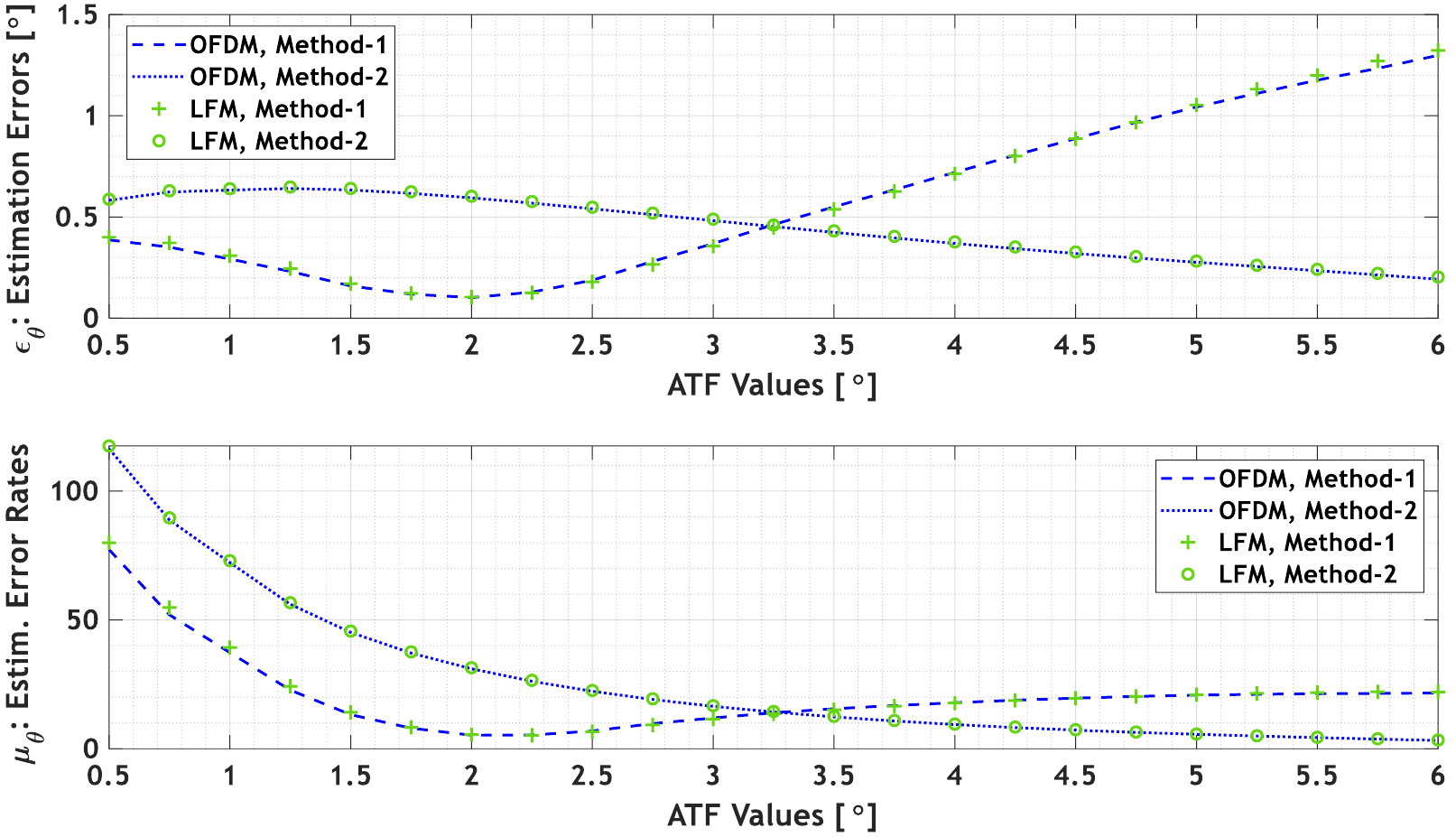}}
    \caption{Estimation accuracy of ATF estimation methods and waveforms over a wider feasible ATF region}
    \label{fig:Figure6}
\end{figure}

\section{Conclusions and Future Works}

In this study we propose an ATF detection method which is followed by two estimation methods to detect and estimate the antenna tilt misalignments exploiting the potential of ISAC technology by treating static environmental objects as fixed anchors and measuring their unexpected shifts in their radar image (CHM). Computer simulations show that both approaches effectively estimate antenna misalignments; however, each method performs best in different scenarios. {\textcolor{black}{Due to space constraints, results are presented for a single transmit power value. While detailed analysis across varying transmit powers-which depend on the radio unit and deployment-is important, the proposed approaches are expected to be robust, as they also include short-range points where the backscattered signal-to-noise ratio (SNR) can be relatively high, supporting reliable ATF detection and estimation.}} With these approaches, 6G networks can self-diagnose physical antenna misalignments, aiding fully autonomous operations. Future work may include enhanced algorithms to improve detection and estimation accuracy or exploiting AI to better distinguish antenna tilts from CHMs{\textcolor{black}{{, especially in highly dynamic environments with significant mobility and under challenging SNR conditions. Under these challenging conditions, a future study providing comprehensive analysis and comparison of the proposed algorithms would be valuable}}}. In addition, testing the proposed system in real-world trials will be of utmost value.

\section*{Acknowledgment}
This study has been supported by the 1515 Frontier Research and Development Laboratories Support Program of T{\"U}B{\.I}TAK under Project 5229901 - 6GEN. Lab: 6G and Artificial Intelligence Laboratory. The authors thank Buse BİLGİN for insightful discussions that contributed to the initial formulation of the ideas presented in this work.



\end{document}